\pdfoutput=1
\documentclass[preprint, amsmath, amssymb, aps, superscriptaddress]{revtex4-2}

\usepackage{graphicx}
\usepackage{dcolumn}
\usepackage{bm}
\usepackage{color}
\usepackage{float}
\usepackage{siunitx}

\begin{document}

\title{Ferrofluid drop impacts and Rosensweig peak formation in a non-uniform magnetic field}

\author{Amelia Cordwell}
\thanks{Joint first author}
\affiliation{Department of Physics, The University of Auckland, Auckland 1142, New Zealand}

\author{Alex Chapple}
\thanks{Joint first author}
\email[\newline Corresponding Author: ]{acha809@aucklanduni.ac.nz} 
\affiliation{Department of Physics, The University of Auckland, Auckland 1142, New Zealand}

\author{Stephen Chung}
\affiliation{Department of Physics, The University of Auckland, Auckland 1142, New Zealand}

\author{Frederick Steven Wells}
\affiliation{Department of Physics, The University of Auckland, Auckland 1142, New Zealand}
\affiliation{The MacDiarmid Institute for Advanced Materials and Nanotechnology, New Zealand}

\author{Geoff R. Willmott}
\email[Corresponding Author: ]{g.willmott@auckland.ac.nz} 
\affiliation{Department of Physics, The University of Auckland, Auckland 1142, New Zealand}
\affiliation{The MacDiarmid Institute for Advanced Materials and Nanotechnology, New Zealand}
\affiliation{School of Chemical Sciences, The University of Auckland, Auckland 1142, New Zealand}

\begin{abstract}
\noindent Vertical drop impacts of ferrofluids onto glass slides in a non-uniform magnetic field have been studied using high-speed photography. Outcomes have been classified based on the motion of the fluid-surface contact lines, and formation of peaks (Rosensweig instabilities) which affect the height of the spreading drop. The largest peaks are nucleated at the edge of a spreading drop, similarly to crown-rim instabilities in drop impacts with conventional fluids, and remain there for an extended time. Impact Weber numbers ranged from 18.0 to 489, and the vertical component of the $B$-field was varied between 0 and 0.37~T at the surface by changing the vertical position of a simple disc magnet placed below the surface. The falling drop was aligned with the vertical cylindrical axis of the 25~mm diameter magnet, and the impacts produced Rosensweig instabilities without splashing. At high magnetic field strengths a stationary ring of ferrofluid forms approximately above the outer edge of the magnet. 
\end{abstract}

\maketitle

\section{\label{sec:level1} Introduction}

Ferrofluids, which consist of ferromagnetic colloids in solution \cite{rosensweig_ferrohydrodynamics_2014,1849}, have found industrial applications such as loudspeaker cooling, sealing rotating shafts, and high-speed printing \cite{kole}. They have been used in areas as diverse as sculpture \cite{kodama_dynamic_2008}, self assembly \cite{timonen_switchable_2013}, medicine \cite{kole}, and even digital logic \cite{katsikis_synchronous_2015}. Ferrofluids are interesting because various complex phenomena are produced when they interact with applied magnetic fields \cite{1849}. Most famously, ferrofluids form Rosensweig instabilities: a series of ordered peaks spontaneously forms on the surface of a ferrofluid in the presence of a magnetic field normal to the surface \cite{rosensweig_ferrohydrodynamics_2014}. After peaks form they slowly arrange into regular hexagonal or square grids, depending on the fluid's magnetisation history \cite{gollwitzer_via_2006}. The creation of such patterns is only possible when the fluid reaches a magnetisation above a critical value as dictated by the fluid parameters \cite{cowley_interfacial_1967}. The heights of Rosensweig peaks increase with increased magnetic field strength \cite{friedrichs_pattern_2001}, although in most cases the parameters of the analytic theory describing this growth have been found empirically \cite{gollwitzer_via_2006}. 

The impact of ferrofluids onto solid surfaces has been the subject of only a few studies for the cases of no or uniform magnetic field \cite{ahmed_maximum_2018,odenbach_colloidal_2009,Reimann_ridges,Jiandong_1,Abrar_2,sahoo_collisional_2021}, even though the general field of liquid droplet impacts is of strong fundamental and applied interest \cite{yarin_review,1475,1131}. It has been found that the spreading of an impacting ferrofluid drop is delayed by a vertical magnetic field, but extends further than with no field applied \cite{ahmed_maximum_2018}. Horizontal magnetic fields have been found to stabilise a ferrofluid surface, delaying the onset of instabilities \cite{sahoo_collisional_2021}. Angled magnetic fields cause the fluid to form ridges rather than hexagonally arranged peaks at high enough field strengths \cite{odenbach_colloidal_2009, Reimann_ridges}. A study reaching relatively high magnetic field strengths (up to $\sim$ 0.1T) showed that the normalized spreading droplet height monotonically decreased with increased magnetic field strength for nanolitre droplet volumes \cite{ahmed_maximum_2018}. Another study \cite{Jiandong_1} at similar field strengths confirmed these findings, and demonstrated that the spreading oscillation of the droplet in the horizontal direction can be reduced substantially with a vertical magnetic field. The maximum spread of impacting ferrofluid droplets has been studied in fields $\lesssim 0.3$~T \cite{ahmed_maximum_2018, Abrar_2}.

It is of interest to study ferrofluid droplet impacts at high enough magnetic field strengths ($\gtrsim 0.15$~T) and with large enough droplet volumes so that Rosensweig peaks play an important role in the spreading dynamics. Moreover it is desirable to understand the dynamics of ferrofluid droplets in non-uniform magnetic fields, for both fundamental and practical reasons. For example, on-demand ferrofluid droplet formations can be created using a time-dependent non-uniform magnetic field \cite{drop_formation_1}. The field generated by a simple bar magnet is also non-uniform. The dynamics of stationary ferrofluid droplets in the presence of a non-uniform magnetic field have been studied \cite{vieu_shape_2018}, but the influence of non-uniform magnetic fields on drop impacts and consequent Rosensweig instability formation is still to be described. A ferrofluid droplet falling prior to impact is elongated in the direction of any applied magnetic field, taking an ellipsoidal shape \cite{shi_experimental_2014}. In a non-uniform field this elongation can produce asymmetries, and an extended tip forms for drops falling vertically along the cylindrical axis of a disc magnet \cite{1848}.

In this work we perform experiments and analysis to study vertical ferrofluid drop impacts in the non-uniform magnetic field generated by a permanent disc magnet placed below the impact surface. Our work extends to comparatively higher magnetic field strengths and impact velocities than previous ferrofluid drop impact studies, and we omit drops which splash or break up. 
We report on the dynamic formation of Rosensweig instabilities following impact. Drop impacts on solid surfaces for any fluid produce a spreading thin lamella, with a raised dynamic rim at the edge of the spreading droplet \cite{Sahoo_rim_formation}. So-called `crown' instabilities on this rim can lead to corona splashes, where secondary droplets are created \cite{yarin_review, yarin_weiss_1995}. For ferrofluids, we are particularly interested in the interplay between the formation of crown instabilities and Rosensweig instabilities, as well as the patterning of these instabilities.

We have also found that at relatively high field strengths, ferrofluid drop impacts can result in stationary rims of ferrofluid forming near the edge of the magnet. Rim formations have been observed previously \cite{Sahoo_rim_formation}, although they were not stationary, likely due to weaker magnetic fields.

\section{Materials and Methods \label{sec:Methods}}

\begin{figure}
    \centering
    \includegraphics[width=8.6cm]{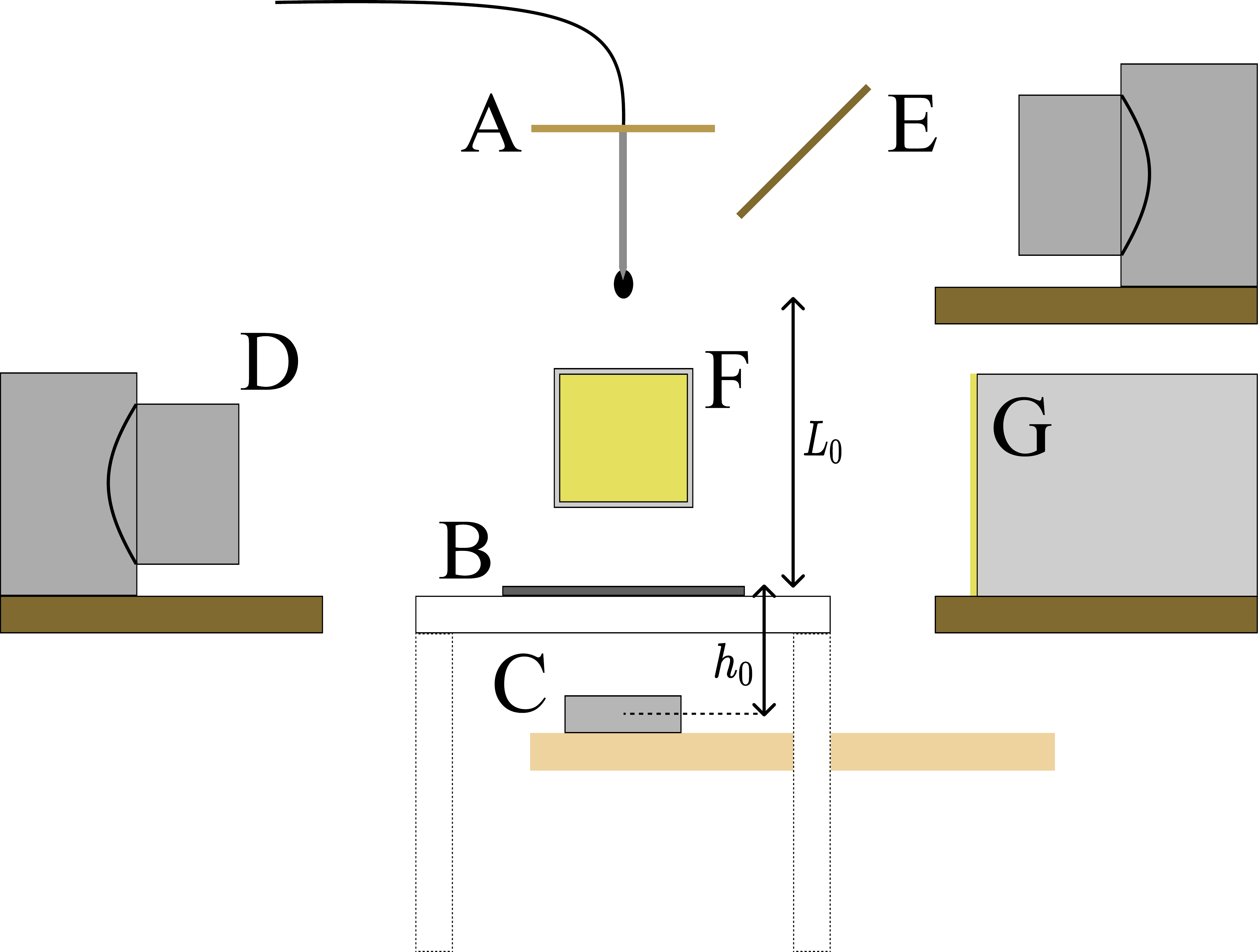}
    \caption{Schematic diagram of the experiment. A is the variable height needle stage, B the glass slide, C the magnet on a movable stage, D the side view camera, E the top view camera with a mirror providing an image at an oblique angle, and F and G are LED lighting sources.} 
    \label{fig:experimentalsetup}
\end{figure}

Ferrofluid droplets were dropped onto a clean glass slide placed above a permanent neodymium disc magnet, and the resulting impacts were studied using high-speed photography (Fig.~\ref{fig:experimentalsetup}). The ferrofluid (EFH1, Ferrotec) was composed of 3 vol\% Fe\textsubscript{3}O\textsubscript{4} particles in light hydrocarbon. The material properties (based on manufacturer's specifications \cite{Ferro_Tec_specs}) include a saturation magnetisation of $M_s = 44$~mT, density $\rho = 1.21 \times 10^3$~kg/m$^3$, initial magnetic susceptibility $\chi = 2.64$, surface tension $\sigma = 0.029$~N/m, and viscosity $\eta = 6$ \si{\milli\pascal \second}.

\begin{figure}
    \centering
    \includegraphics[width=8.6cm]{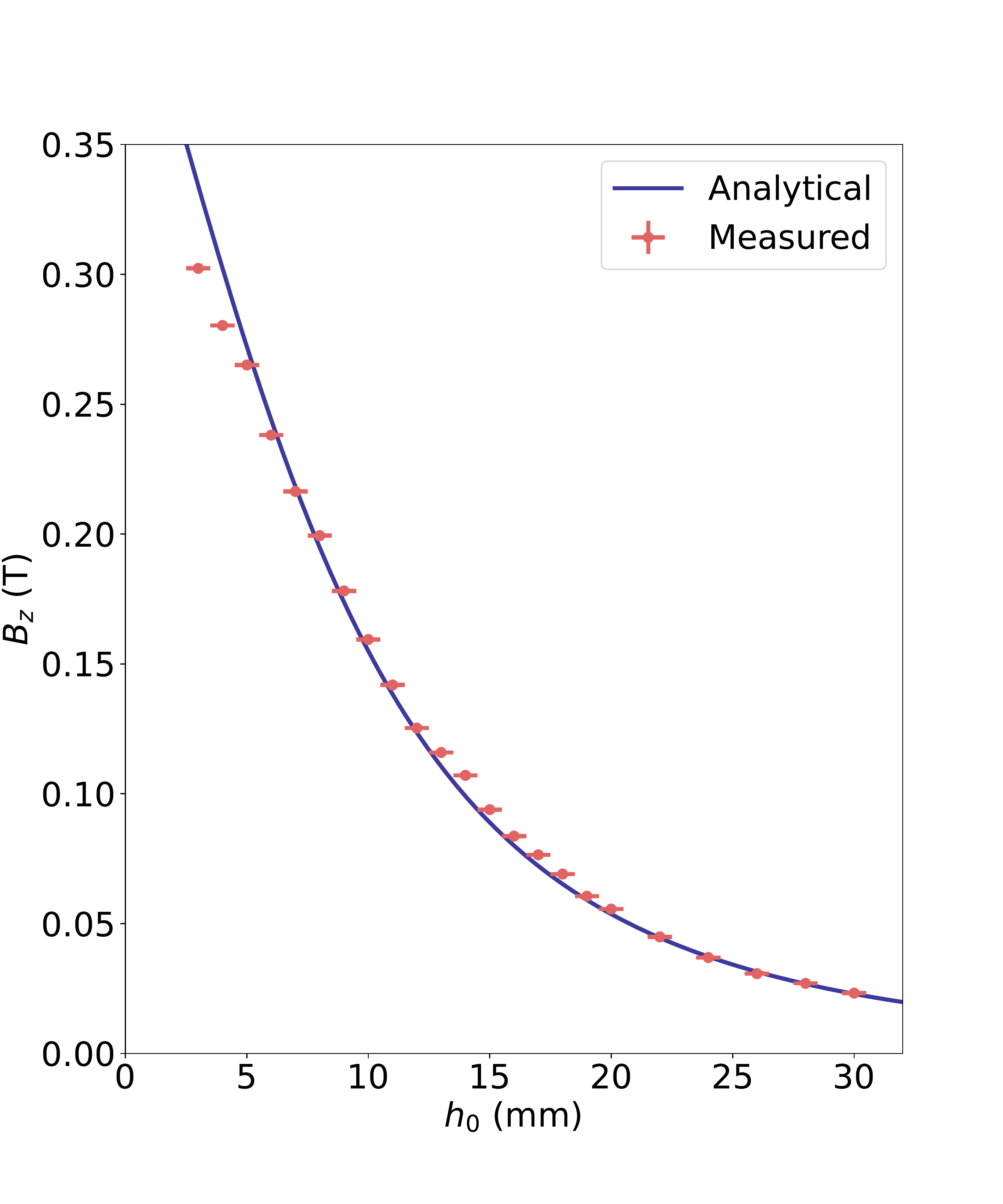}
    \caption{Measured (points) and modelled (line) on-axis vertical magnetic flux density for the 25~mm diameter disc magnet used in this study, plotted as a function of the magnet-surface separation $h_0$. The model line represent a uniformly magnetised cylinder with on-axis magnetisation of $M = 1.22/\mu_0$~T.}
    \label{fig:mag_field_strength}
\end{figure}

The magnet was an axially magnetised cylindrical disc (NdFeB Grade-N38, AMF Magnetics Australia) with diameter of $d = 25$~mm and a length of $l = 12.7$~mm. It was affixed to a movable platform so that the distance between the top of the magnet and the top of the glass slide ($h_{0}$) could be varied. The vertical component of the magnetic flux density ($B_z$) was measured using a teslameter (Group 3 DTM-133) at radial positions of $R=0, 6.2$ and $12.5$~mm for cylindrical polar co-ordinates $(R, \phi, z)$ with their origin at the centre of the upper surface of the magnet. The resulting plot of $B_z$ as a function of $h_{0}$ (equivalent to $z$) is shown in Fig.~\ref{fig:mag_field_strength}. A numerical model \cite{Blinder_cylinder_magnet} for a uniformly magnetised cylinder of the same size agreed with the on-axis measurements for a magnetisation of $M = 1.22/\mu_0$~T (here $\mu_0$ is the permeability of free space). The accuracy of the model was reduced off-axis, likely because the magnet is not exactly uniformly magnetised. Repeat measurements carried out using the same magnet over a long time span ($\sim$3 years) have revealed no significant change in the on-axis field strength. It was found that the critical value of $h_0$ for Rosensweig peak formation for a ferrofluid drop on the glass slide was $\approx40$~mm, corresponding to an on-axis vertical flux density of around 12~mT.

Individual droplets were generated by a syringe pump and dropped from a polyether ether ketone (PEEK) needle mounted on a movable stage. The release point was at a variable height $L_0$ above the glass slide surface. The size of the droplets could vary depending on the precise positioning of the needle and on the magnet position, as the droplet release could be affected by the $B$-field. Furthermore, ferrofluid droplets elongate in the direction of the $B$-field resulting in a dynamically varying shape as the droplet falls \cite{1848}. Initial tests were performed to align the centre of the magnet with the centre of the droplet impact. As the paramagnetic ferrofluid is attracted to areas of higher magnetic field, the centre of a misaligned drop would shift towards the center of the magnet approximately a minute after impact, leaving behind some residue. The centre of the magnet could be shifted appropriately. In impact videos, significantly misaligned droplets could be identified due to the non-vertical drop elongation. These were excluded from analysis of height and line dynamics (Fig.~\ref{fig:contact_line_phase}) but some asymmetric impacts are included and discussed in the study of rim formation (Fig.~\ref{fig:rim_phase}). 

Droplet impacts were recorded for magnet distances ranging from $h_0=2$~mm to 28~mm, with drop heights ranging from $L_0 = 260$~mm to 820~mm. The glass slide was cleaned with acetone and wiped with a paper towel in between each impact. Two high speed cameras (a Photron Fastcam SA5 and a Photron Fastcam Mini AX5) operating at 2000~fps (frames per second) were used to record the impacts from side-on and top-down views respectively. The top down view was reflected from a mirror at $\sim 45^{\circ}$ to the vertical, although a more oblique angle was sometimes used (Fig.~\ref{fig:experimentalsetup}). Top-down views were not used to make quantitative measurements, and were analysed manually where required. For the side-view camera, impacts were back-lit using a customised LED lighting unit, and images were spatially calibrated (typically at $\sim$25~$\mu$m per pixel) using a fiducial reference square. 100 frames (0.05~s) were recorded for each impact, long enough to show the initial formation of Rosensweig peaks. In some experiments, recording continued for a longer time period to capture the droplet settling further.

\section{Analysis}

\subsection{Image Analysis and Dimensionless Numbers}

Drop impacts are most often characterised using the Weber number, 

\begin{equation}
    We_0 = \frac{\rho v^2 D_0}{\sigma},
    \label{eq:We0}
\end{equation}

\noindent due to the importance of competition between inertia and surface tension in the drop outcome. Here $D_0$ is the diameter of the droplet, assumed to be spherical, and $v$ is the impact velocity. However, ferrofluid drops in a field become non-spherical, so in this study a first-principles definition of the Weber number is used,  

\begin{equation}
    We = 12 \frac{E_{\text{kin}}}{E_{\text{surf}}} = \frac{6 mv_c^2}{\sigma A_s}, \label{eqn:We_first_principles}
\end{equation}

\noindent where $E_{\text{kin}} = \frac12 mv_c^2$ and $E_{\text{surf}} = \sigma A_s$ are respectively the kinetic and surface energies of the drop, and where $m$ and $A_s$ are respectively the drop's mass and surface area. Furthermore, different parts of the droplet have different velocities at impact due to drop deformation in the non-uniform magnetic field \cite{1848}. Therefore, the center-of-mass velocity $v_c$ is used as the velocity scale. 

Measurements of the drop's initial geometry and velocity were made using the frame immediately prior to impact captured by the side-view camera. Codes were developed in Python to calculate each droplet's volume ($V$) and surface area ($A_s$) by treating each row of pixels as a flat truncated cone one pixel in height, and with the bottom and top radii calculated using the respective widths of the imaged droplet. The measured volume was used to calculate an equivalent diameter $D_0$ for each droplet, defined as the diameter of a spherical droplet with same volume i.e. $D_0 = 2 (3 V$/$4 \pi )^{1/3}$. Analysis of the experimental images revealed that $v_c$ varied between $0.50$ and $3.88$~m/s. The range of equivalent diameters was $0.45 < D_0 < 11.6$~mm, and the experiments covered the range $18.0 < We < 489$.

The value of $We$ calculated using Eq.~\ref{eqn:We_first_principles} is the dimensionless number primarily used to describe the experiments here. A comparison with other dimensionless numbers is provided in the Supplemental Material \cite{APS_SI}, using the $D_0$ and $v_c$ as length and velocity scales respectively. These include $We_0$ (Eq.~\ref{eq:We0}), as well as the Reynolds ($Re = \rho v_c D_0$/$\eta$), magnetic Bond ($Bo_m = B_z^2 D_0$/$\mu_0 \sigma$) and Hartmann ($Ha = \rho B \mathcal{M} D_0$/$2 v_c \eta$, where $\mathcal{M}$ is the magnetic moment) numbers. These dimensionless numbers are not used in the main text because the use of $D_0$ (unlike the calculation of $We$) does not account for droplet elongation.

To study the drop spreading dynamics, the contact line width ($s$, see Supplemental Material \cite{APS_SI}) and the maximum height of the droplet above the surface ($H$) were also measured as a function of time after impact ($t$). Drop impact studies often use a dimensionless spread factor, defined as the linear extent of the spreading drop divided by the drop's diameter prior to impact. Here the droplets can be highly non-spherical prior to impact, so an alternative spread factor ($\beta$) is used, in which the contact line width is divided by the equivalent diameter,

\begin{equation}\label{eq:spread}
    \beta = \frac{s}{D_{0}}.
\end{equation}

\noindent The height is similarly non-dimensionalized using $\delta = H$~/~$D_0$.

Measurements taken at the point of maximum spreading can be used for comparisons of drop impacts. However, it was found that for a ferrofluid drop in a magnetic field the contact line width can reach an initial maximum, then steadily grow or else decrease before growing again due to magnetic effects (see below). Therefore we identify maximum spread with the first maximum of the contact width after impact.  

\subsection{The Radial Field}\label{Sec:rad}

A theoretical magnetic body force acting on the fluid in a radial direction can be calculated. This force is caused by the radial change in the magnetic field  magnitude. For this calculation it is assumed that the fluid is at saturation magnetisation, that the force on any volume of fluid can be considered independently, and that the magnet is a cylinder of axially uniform magnetisation. Under these conditions, the force on a magnetised material in an external field,

\begin{align}
     \mathbf{F} = \int_V \nabla (\mathbf{M} \cdot \mathbf{B}) dV, 
\end{align}

\noindent produces a magnetic body force on a unit volume of ferrofluid at saturation magnetisation, $\mathbf{M} = M_s \mathbf{B}/|B|$, 

\begin{align}
    \mathbf{f} = M_s\nabla(\mathbf{B}/|B| \cdot \mathbf{B}),
\end{align}

\noindent with the radial component

\begin{align}\label{Eqmag}
    f_{R} = M_s \frac{\partial }{\partial R} (\mathbf{B} \cdot \mathbf{B})^{1/2}. 
\end{align}

\begin{figure}
    \centering
    \includegraphics[width=8.6cm]{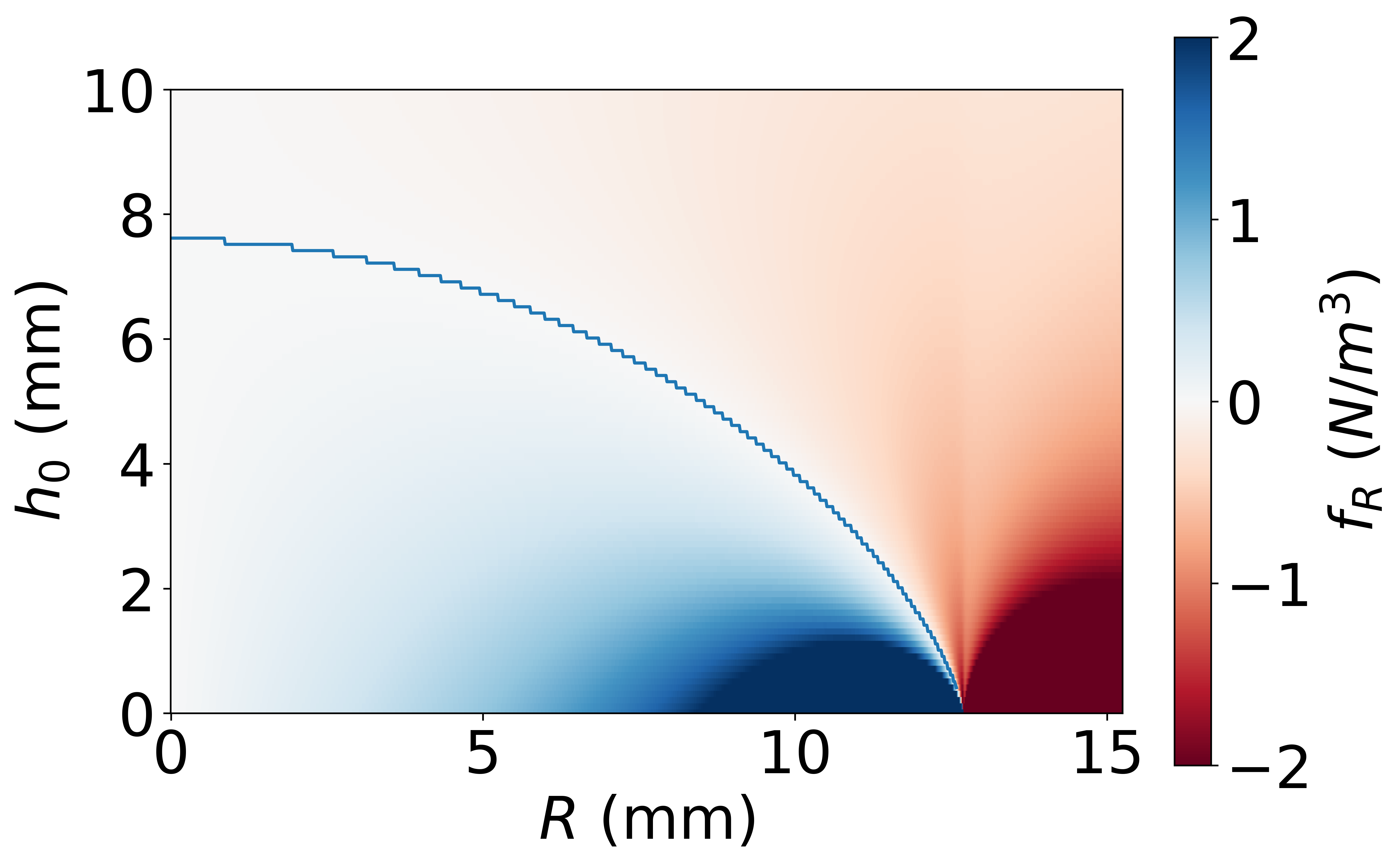}
    \caption{Radial magnetic force per unit volume of ferrofluid at the impact surface for an axially magnetised uniform cylinder as described in the text. The blue line corresponds to $f_R=0$.}
    \label{fig:rim_forces}
\end{figure}

Figure \ref{fig:rim_forces} shows the radial force per unit volume of ferrofluid as a function of $R$ at the impact surface (a distance $h_0$ above the magnet), computed using Eq.~\ref{Eqmag} and the $B$-field generated from the fit to the data in Fig.~\ref{fig:mag_field_strength}. When the surface is more than 7.5~mm above the magnet, there is no radial outwards force ($f_R<0$ for all values of $R$). When $h_0<7.5$, an outward radial force acts at radial positions from the centre of the magnet ($R=0$) out to a point where the sign of $f_R$ changes. As $h_0$ decreases the radial position at which $f_R=0$ increases, and elsewhere the magnitude of $f_R$ increases. The signs of the force indicate stable equilibrium at $f_R=0$, and when $h_0$ is small the restoring forces intensify. This calculation suggests that ferrofluid could be pushed towards $f_R=0$, potentially forming a circular rim which would become thinner and further from the centre of the magnet with decreasing $h_0$. 

\section{Results and Discussion}\label{Sec:ros}

\subsection{Typical Results}\label{Sec:typ}

\begin{figure}
    \centering
    \includegraphics[width=8.6cm]{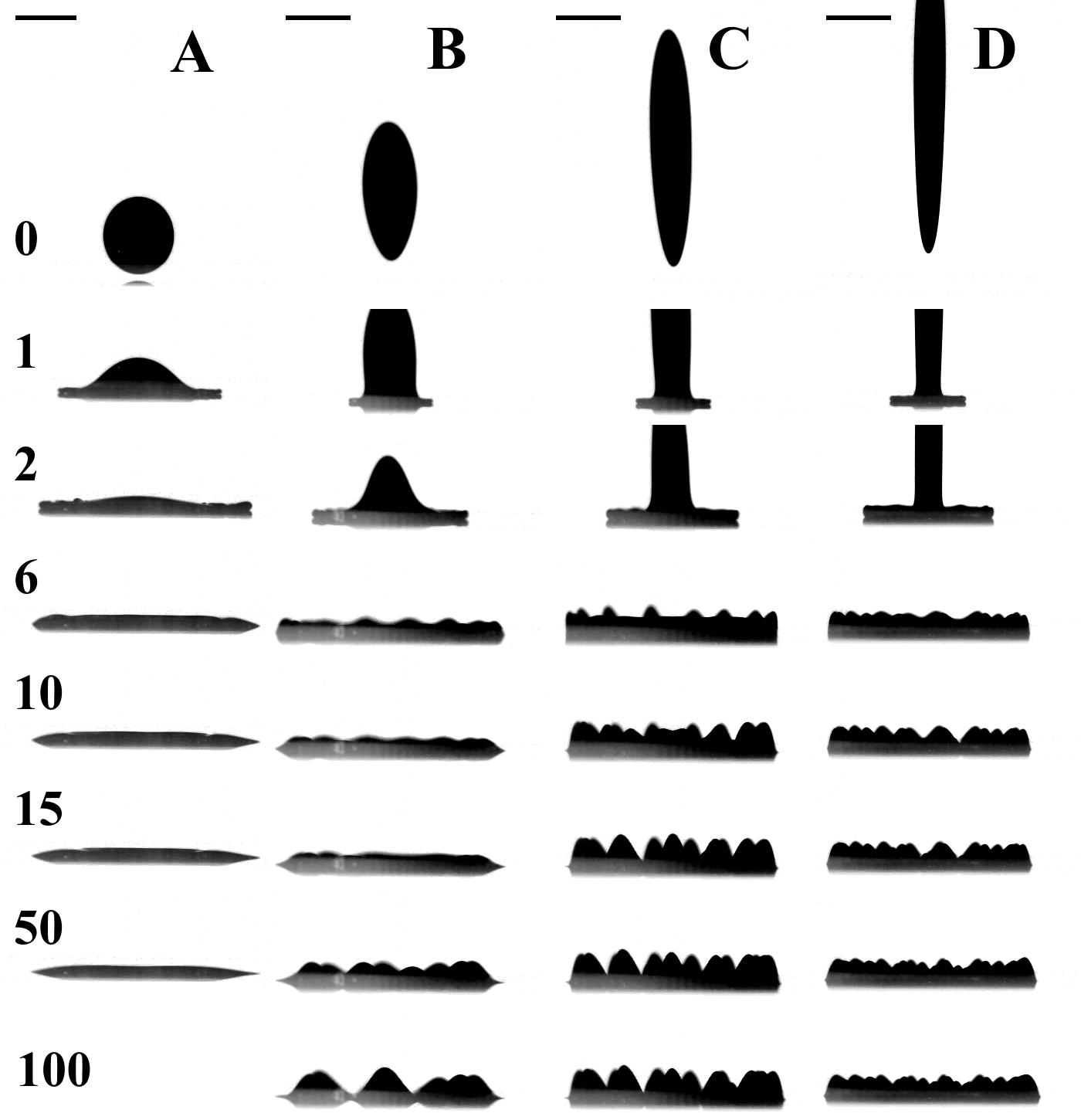}
    \caption{Droplet impact sequences in vertical strips, labelled with the time after impact in milliseconds. Scale bars are 1~mm.
    Droplet A: We = 289, $V = 7.34$~$\mu$l, $v = 1.61$~m/s, $h_0 = 62.0$~mm, $B_z = 0.007$~T.
    Droplet B: We = 271, $V = 6.25$~$\mu$l, $v = 1.72$~m/s, $h_0 = 28.9$~mm, $B_z = 0.025$~T.
    Droplet C: We = 249, $V = 6.36$~$\mu$l, $v = 1.82$~m/s, $h_0 = 12.85$~mm, $B_z = 0.112$~T.
    Droplet D: We = 269, $V = 4.82$~$\mu$l, $v = 2.14$~m/s, $h_0 = 5.0$~mm, $B_z = 0.272$~T.}
    \label{fig:drop_video}
\end{figure}

\begin{figure}
    \includegraphics[width=8.6cm]{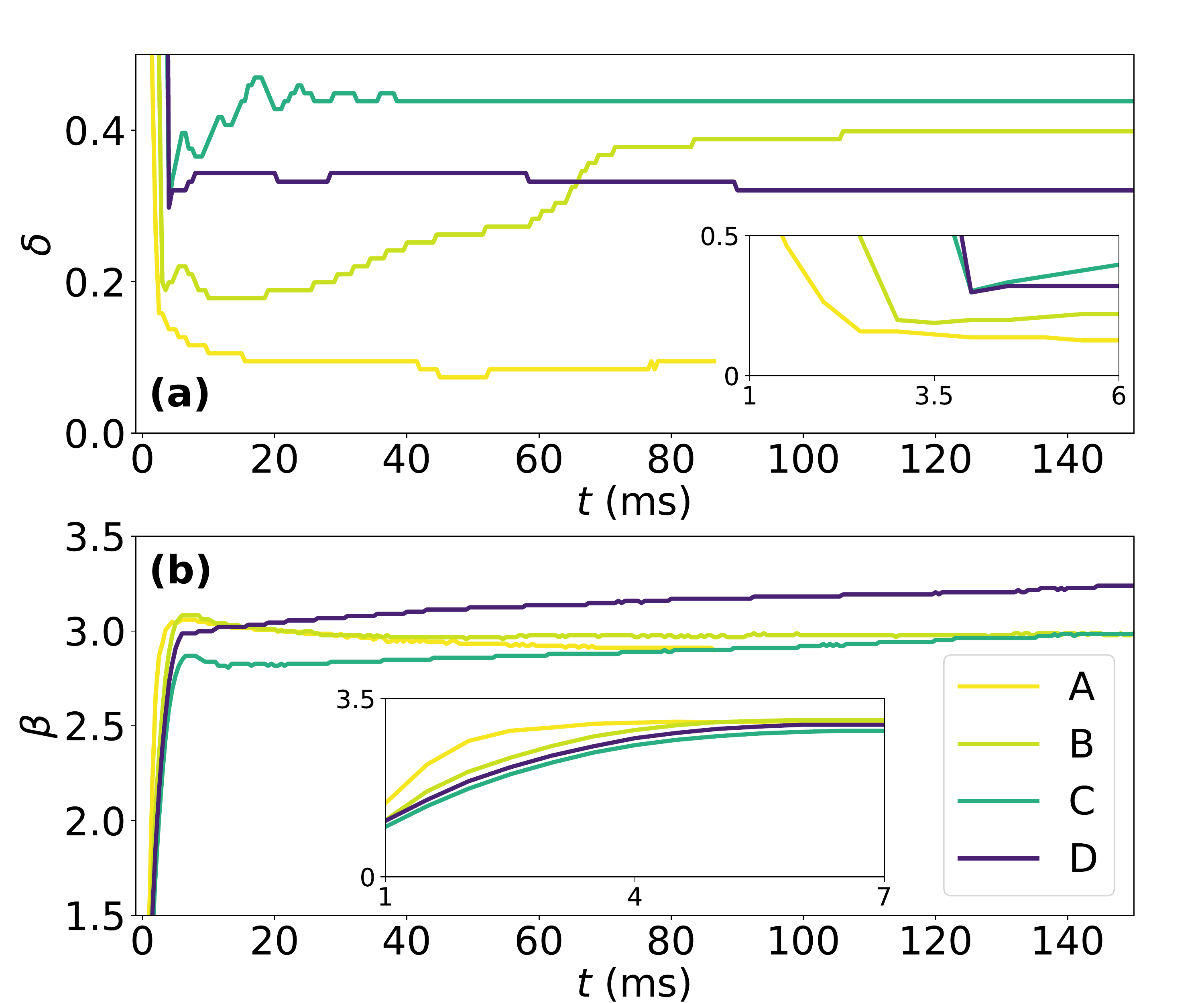}
    \caption{For the four droplets (A-D) from Fig.~\ref{fig:drop_video}, plots of (a) normalized height of the droplet shape, and (b) spread factor as defined in Eq.~\ref{eq:spread}. Insets show an expanded view of the first few ms for each figure.}
    \label{fig:height_contact_example}
\end{figure}

Figure \ref{fig:drop_video} shows four representative examples of ferrofluid droplet impacts from this study, with their height and contact line dynamics recorded in Fig.~\ref{fig:height_contact_example}. Full videos of impacts A-D are contained in the Supplemental Material \cite{APS_SI}. The magnetic field increases monotonically from A to D, affecting the elongated shape at impact \cite{1848}, and the centre-of-mass velocity at impact progressively increases. The drop size decreases with $h_0$ because the field also affects drop release. Droplet A is not under significant influence of the magnetic field and simply spreads out before retracting slightly. This deposition outcome is similar to observations of ferrofluid drop impacts onto hydrophilic surfaces with no applied field reported previously \cite{sahoo_collisional_2021}. Droplets B-D have different dynamics. 
For droplet B the initial impact and spreading produce a height and spread factor similar to droplet A (Fig.~\ref{fig:height_contact_example}), despite the elongation of the drop at impact. In general, the maximum spread factor observed in this study at relatively low field strengths ($B_z\lesssim100$~mT) follows a trend close to $\beta_{max}\propto We^{0.25}$ (see Supplemental Material \cite{APS_SI}), as is often observed for non-magnetic drop impacts onto solid surfaces \cite{Marengo}. Unlike droplet A however, instabilities appear on the outer edge of the rim of droplet B as it approaches maximum spread ($\approx 6$~ms), similar in appearance to crown-rim instabilities observed for conventional drop impacts \cite{yarin_review, yarin_weiss_1995}. These instabilities do not generate secondary drops or large fingers, but as the droplet retracts, they do not disappear as they would for a conventional drop impact. Between 15 and 50~ms, the growth of Rosensweig instabilities becomes apparent. Qualitatively, we observe that there appears to be a correlation between the crown-like instabilities and the Rosensweig instabilities. This suggests a mechanism in which the Rosensweig peaks nucleate from the rim instabilities, which is reasonable given that Rosensweig peaks are caused by a relative increase in the magnetic force acting on small peaks on an otherwise uniform film. Once nucleated, peaks undergo unstable growth until opposed by sufficient surface tension \cite{rosensweig_ferrohydrodynamics_2014}. 

For droplet C, fluctuations in the rim when the drop is near maximum spread ($\approx 6$~ms) are directly observed to evolve into Rosensweig instabilities. Near maximum spread, the peaks emerging from the expanding lamella are more pronounced than for either droplet B or D. Over time, some of the instabilities shift and merge into one another (see videos in the Supplemental Material \cite{APS_SI} and Fig.~\ref{fig:vid_topdown}, below), and consequently the maximum height appears to fluctuate. These movements reflect that the spatial arrangement of crown-rim instabilities on a dynamically moving lamella does not correspond to a stable (or metastable) arrangement of Rosensweig peaks. The number of peaks and the droplet height remain constant after $\approx 40$~ms, but the peaks continue to spread apart from each other. In the contact line dynamics (Fig.~\ref{fig:height_contact_example}(b)), after maximum spread the drop initially retracts slightly as for droplets A and B. Later droplet C spreads out, which is discussed in Section~\ref{Sec:class}. Droplet C produces shapes with the greatest minimum height and lowest maximum spreading of the four droplets (Fig.~\ref{fig:height_contact_example}). The height is produced by the clear and early emergence of peaks, while it is generally observed that the maximum spread reduces with $B_z$ at relatively low fields (see Supplemental Material \cite{APS_SI}).

For droplet D, Rosensweig instabilities form earlier than for droplet C, and the contact line moves continuously outwards without retracting. The peak spacing decreases monotonically between droplets B, C and D. The maximum Rosensweig peak width in vertically non-uniform fields can be calculated as \cite{timonen_switchable_2013, vieu_shape_2018},

\begin{equation}
    \lambda_c = 2\pi \sqrt{\frac{\sigma}{\rho g + M_s\frac{d}{dz}(B_z)}}, \label{eqn:peak_spacing}
\end{equation}

\noindent where $g$ is gravitational acceleration, and it is assumed that the magnetic flux density is only dependent on $z$ and that the fluid is at its saturation magnetisation. It is therefore reasonable to attribute the decreasing peak spacing for droplets B-D to the increasing magnetic field gradient. 

We note two questions that are apparent when comparing droplets A-D that are left as suitable topics for future study. Firstly it is observed that rim instabilities evolve into Rosensweig peaks, but it is unclear whether the origin of the rim instabilities is always magnetic. A comparison between droplets A and B suggests this is the case, but it is possible that the shape and velocity of the impact for droplet B would generate rim instabilities in the absence of a $B$-field. Here we note that the time scale for Rosensweig instability growth decreases with increasing magnetic field, so that the effects of magnetism are clear at maximum spread for droplets C and D. Secondly the increasing aspect ratio of the droplets before impact may have an important influence on the dynamics, as the upper halves of these droplets experience conditions similar to that of a conventional droplet impacting on a thin film. A recent study has approached the topic of elongated drop impacts, albeit for regular ellipsoids with aspect ratios 2:1 or smaller \cite{aspect_ratio}. That study found that the scaling of maximum spreading diameter with We$^{0.25}$, as developed for spherical droplets, is preserved. Here, we do not attempt to deconvolute the separate influences of impacting droplet shape and the magnetic field on the spreading lamella.

\parskip=0pt

\subsection{Impact Outcome Classification}\label{Sec:class}

Droplets outcomes have been classified for a large number of experiments using measurements such as those in Fig.~\ref{fig:height_contact_example}. Experiments were focused at $We$ values and release heights below those that resulted in the droplet breaking up or splashing, and any such cases were identified and excluded. The height dynamics were divided into two categories based on the trend observed following the initial spreading (up to maximum spread, or $\approx$5~ms) of the droplet. These are heights that monotonically decrease (Decay, e.g. droplet A), or else increase (Grow, e.g. droplets B and C). The contact line dynamics were classified into three categories, also based on the trend observed following the initial spreading of the droplet. Decay occurs when $\beta$ increases to a maximum spread before decreasing (e.g. droplets A and B), Rebound occurs when $\beta$ initially decreases after reaching maximum spread before eventually increasing above that value (e.g. droplet C), and Rise occurs when $\beta$ monotonically increases (e.g. droplet D). 

\begin{figure}
    \centering
    \includegraphics[width=8.6cm]{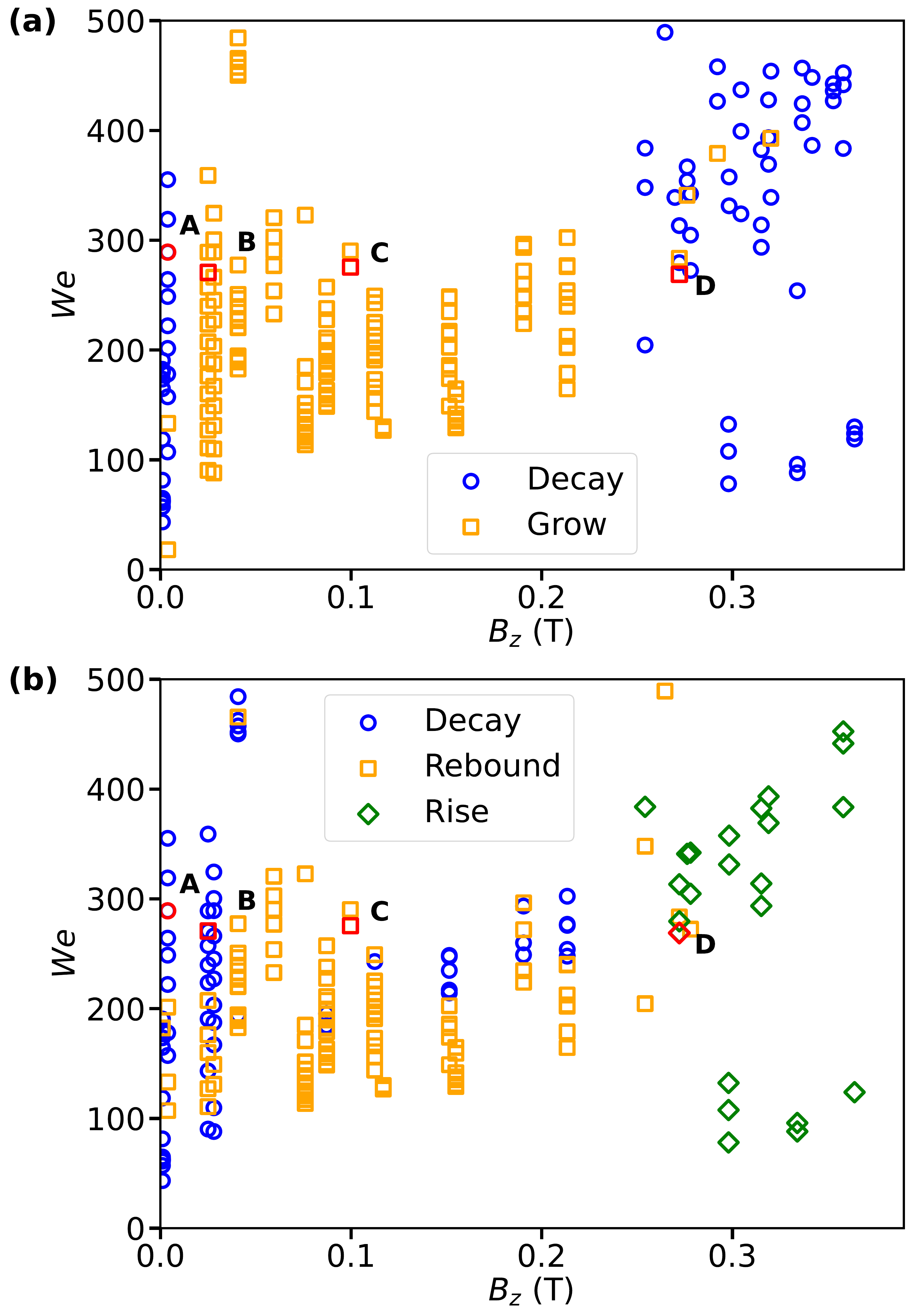}
    \caption{For ferrofluid drop impacts, classifications of (a) height and (b) contact line dynamics as discussed in the text. Red markers indicate the positions of experiments A - D from Figure \ref{fig:drop_video}, as labelled. Data points in (a) and (b) do not exactly correspond because the imaging quality precluded a classification in some cases.}
    \label{fig:contact_line_phase}
\end{figure}

Plots of outcomes as a function of $B_z$ and $We$ shown in Fig.~\ref{fig:contact_line_phase} give generalised evidence for the trends identified in Section~\ref{Sec:typ}. The trend for droplet height (Fig.~\ref{fig:contact_line_phase}(a)) is notable because it is non-monotonic with respect to field strength. The Decay classification is recorded when there is no magnetic field present, as expected for a non-magnetic fluid on a wetting surface. When a relatively small magnetic field is present, emergence of Rosensweig peaks leads to an increase in the height (Grow classification), as expected from previous observations \cite{friedrichs_pattern_2001}. For $B \gtrsim 0.25$~T the Decay classification is again recorded because the Rosensweig peaks are small and form almost immediately. The classifications can be uncertain (especially near boundaries) because the degree of height increase or decrease can be small. For example for droplet D, the height remains approximately constant (Fig.~\ref{fig:height_contact_example}(a); classified marginally as Grow), consistent with its position on the boundary between Decay and Grow classifications in Fig.~\ref{fig:contact_line_phase}(a). 

The contact line dynamics in Fig.~\ref{fig:contact_line_phase}(b) broadly indicate a transition from Decay to Rebound to Rise as the magnetic field strength increases. At zero field the drops retract slightly due to the wetting properties of the fluid following impact. With a field applied, two apparently contradictory trends are observed: the maximum spread following impact decreases with $B_z$ (see Supplemental Material \cite{APS_SI}), while later the spreading tends to increase with $B_z$, producing Rebound and Rise classifications. The former trend (but not the latter) is supported by the calculation of radial force components (Fig.~\ref{fig:rim_forces}), which suggests that $f_R$ always acts inwards for $B_z \lesssim 0.20$~T ($h_0 \gtrsim 7.5$~mm). The outward droplet spreading can be instead explained by the Rosensweig peaks, which repel each other, and can therefore force the (wetting) contact line to gradually spread. This mechanism is consistent with the apparent dependence of the classifications on $B_z$ (and hence on the vertical field which pins the peaks) in Fig.~\ref{fig:contact_line_phase}(b) and with the correspondence in timing and intensity between peak formation and contact line growth in Fig.~\ref{fig:height_contact_example}(b).

To summarize Fig.~\ref{fig:contact_line_phase}(b), at low non-zero fields ($B_z \lesssim 0.04$), the field has insufficient strength to reverse the retraction observed at zero field. A few relatively tall Rosensweig peaks form well after the initial spreading of the droplet, at which point $\beta$ may increase relative to the zero-field case (Fig.~\ref{fig:height_contact_example}(b)). Over a wide range of higher field strengths ($0.04 \lesssim B_z \lesssim 0.25$), the increased pinning of the peaks produces a spreading force that prevails following an initial small retraction, so that the contact line width increases steadily in the long time limit. In this range, Decay outcomes are observed for some cases at relatively high $We$, in which the spreading and retraction driven by inertia dominates any spreading caused by peaks. Finally, for the highest fields studied ($B_z \gtrsim 0.25$) a strong outward force pulls the droplet outwards continuously post-impact. Here, outwards motion may be assisted by the radial force (Fig.~\ref{fig:rim_forces}). These Rise outcomes can lead to formation of an outer rim, a phenomenon discussed in Section~\ref{Sec:rim}.

Overall, Fig.~\ref{fig:contact_line_phase} indicates strong correspondence between the transition from Decay to Grow in the height dynamics, and from Rebound to Rise in the contact line dynamics ($B_z \approx 0.25$). The categories are not perfectly delineated as a function of $B_z$ and $We$, highlighting the difficulty of summarising these nuanced experiments using a single dimensionless quantity as discussed above. Differences in droplet volume and in the precise alignment of the droplet impact point with the magnet centre are likely explanations for the indistinct classification boundaries.

\subsection{Motion of Peaks}

\begin{figure}
    \centering
    \includegraphics[width=8.6cm]{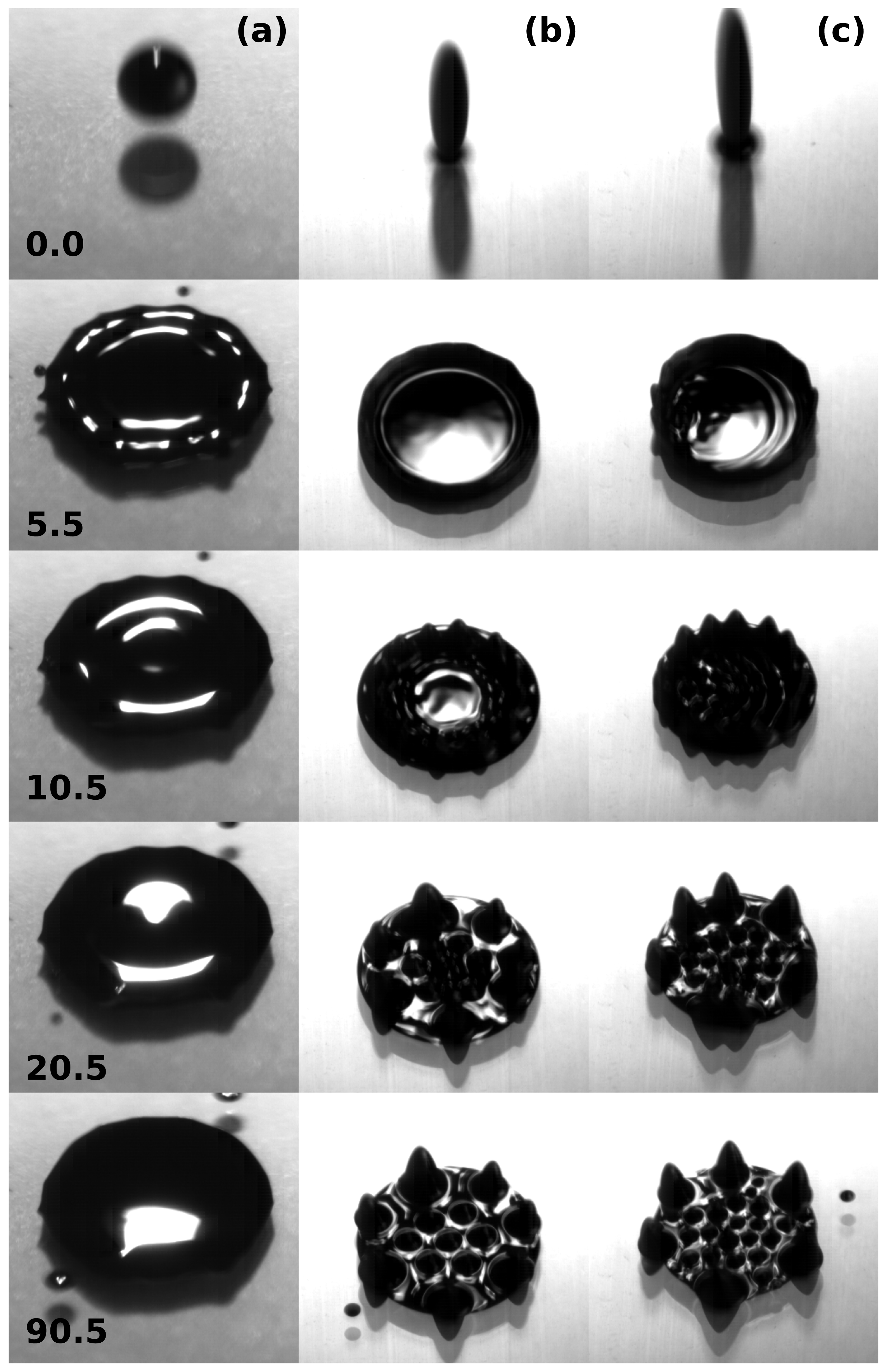}
    \caption{Oblique top-down view of two ferrofluid drop impacts. Experimental parameters are (b) $We$ = 220, $V = 5.05$~$\mu$l, $v=2.00$~m/s, $h_0 = 23$~mm, $B_z=0.041$~T, and (c) $We$ = 322, $V = 4.66$~$\mu$l, $v=2.27$~m/s, $h_0 = 16.5$~mm, $B_z=0.076$~T.}
    \label{fig:vid_topdown}
\end{figure}

Figure~\ref{fig:vid_topdown} uses oblique top-down views of three droplets to explore how the arrangement of the Rosensweig peaks is affected by the dynamics of the drop impact. The drops in the non-zero field (Figs.~\ref{fig:vid_topdown}(b) and (c)) are both classified as ``Grow'' for their height dynamics and ``Rebound'' for their contact line dynamics. 

From impact up to $\approx$10~ms later, the lamella rim for the droplets in a field is more prominent that for the control experiment (Fig.~\ref{fig:vid_topdown}(a)), and instabilities start to become apparent. Peaks form on the spreading rim of the lamella, and are more widely spaced at lower field strength (Fig.~\ref{fig:vid_topdown}(b)). Peaks inside the rim form later and are clearly smaller for two reasons. Firstly, during the spreading phase the thick rim supporting the growth of the outer peaks contains more fluid than the thinner internal lamella. Secondly, the instabilities in the rim form first, during spreading, potentially nucleated by crown-rim instabilities (see Fig.~\ref{fig:drop_video}). These peaks then have longer to grow (and perhaps merge) to create taller peaks than peaks growing from the lamella. The volume of fluid in each peak is less than would be supported by a thick pool of fluid \cite{friedrichs_pattern_2001}, so it is likely that growth of the inner peaks is restricted by a shortage of fluid to draw upon. 

Images captured approximately 1 minute after impact confirmed that, although there was some movement of the Rosensweig peaks, configurations such as those observed after 90~ms in Fig.~\ref{fig:vid_topdown} were reasonably stable. The peaks repel each other and apparently reach a stable state in which fluid is not exchanged between them. The inner peaks do not undergo sufficient merging to grow in size significantly. The observation of larger outer and smaller inner peaks is the inverse of the trend observed for a static droplet on a dry plate, in which case the central peak is significantly larger \cite{chen_breakup_2006}. 

\subsection{Rim Formation}\label{Sec:rim}


\newcommand{\RomanNumeralCaps}[1]{\MakeUppercase{\romannumeral #1}}

\begin{figure}
    \centering
    \includegraphics[width=8.6cm]{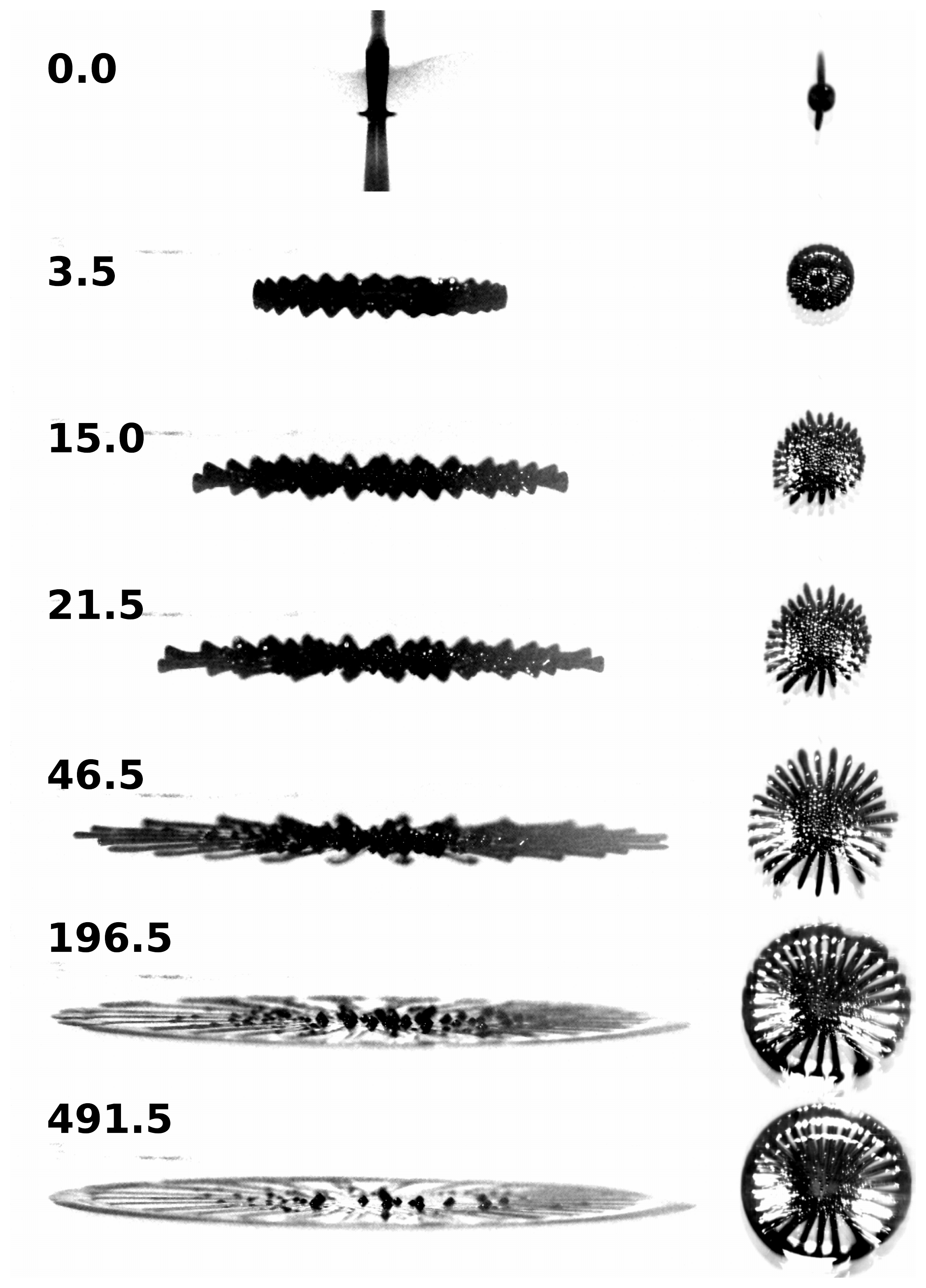}
    \caption{Side-on (left) and top-down (right) oblique views of a ferrofluid drop impact on to a plastic Petri dish with parameters $h_0$ = 1.2~mm, $V\approx 6.2~\mu$l, $L_0$ = 290~mm. Labels denote time in ms.
    }
    \label{fig:rim_formation}
\end{figure}

In Fig.~\ref{fig:contact_line_phase}(b), it was established that the contact line continues to spread outwards after the initial $\approx$5~ms following impact if the magnetic field is strong enough (i.e. the Rise classification). To extend the study of this spreading, some separate experiments were carried out with the magnet placed very close to the surface ($h_0$ = 1.2~mm). A plastic Petri dish was used due to the extent of the spreading drop, and so data from these experiments were not included with the data for drops onto glass in the other figures. Other experimental parameters are as described in Section \ref{sec:Methods}.

Figure \ref{fig:rim_formation} shows a typical example of these impacts. Rosensweig peaks form on the edge of the droplet almost immediately upon impact, and maintain their shape as they move radially outwards. As the peaks spread, they each leave behind a trail of fluid, and their height decreases. Other peaks form nearer the point of impact, and move outwards more slowly. 

Eventually ($\approx 200$~ms) the outward motion is halted and the fluid at the edge of the drop forms a circular rim, approximately following the outline of the edge of the cylindrical magnet below the surface. This outer rim is stationary, without any Rosensweig peaks, and is not to be confused with the rim observed at the edge of the lamella immediately after impact. Well after the rim has stabilised, the slow-moving peaks formed near the impact point can move towards and coalesce with the rim, if the magnetic field is strong enough (see also Supplemental Material \cite{APS_SI}). 

A loss of peak height in spreading ferrofluid drops has previously been attributed to a reduction in vertical magnetic field strength \cite{odenbach_colloidal_2009}, but here the loss in height is likely to be due to increased magnetic field gradient and stabilisation of the rim due to the radial field component. Rim formation in a ferrofluid droplet has been observed previously \cite{Sahoo_rim_formation}, but this occurred in a uniform vertical magnetic field created by electromagnetic coils, and the rim was not pinned as in the present experiments.  


\begin{figure}
    \centering
    \includegraphics[width=8.6cm]{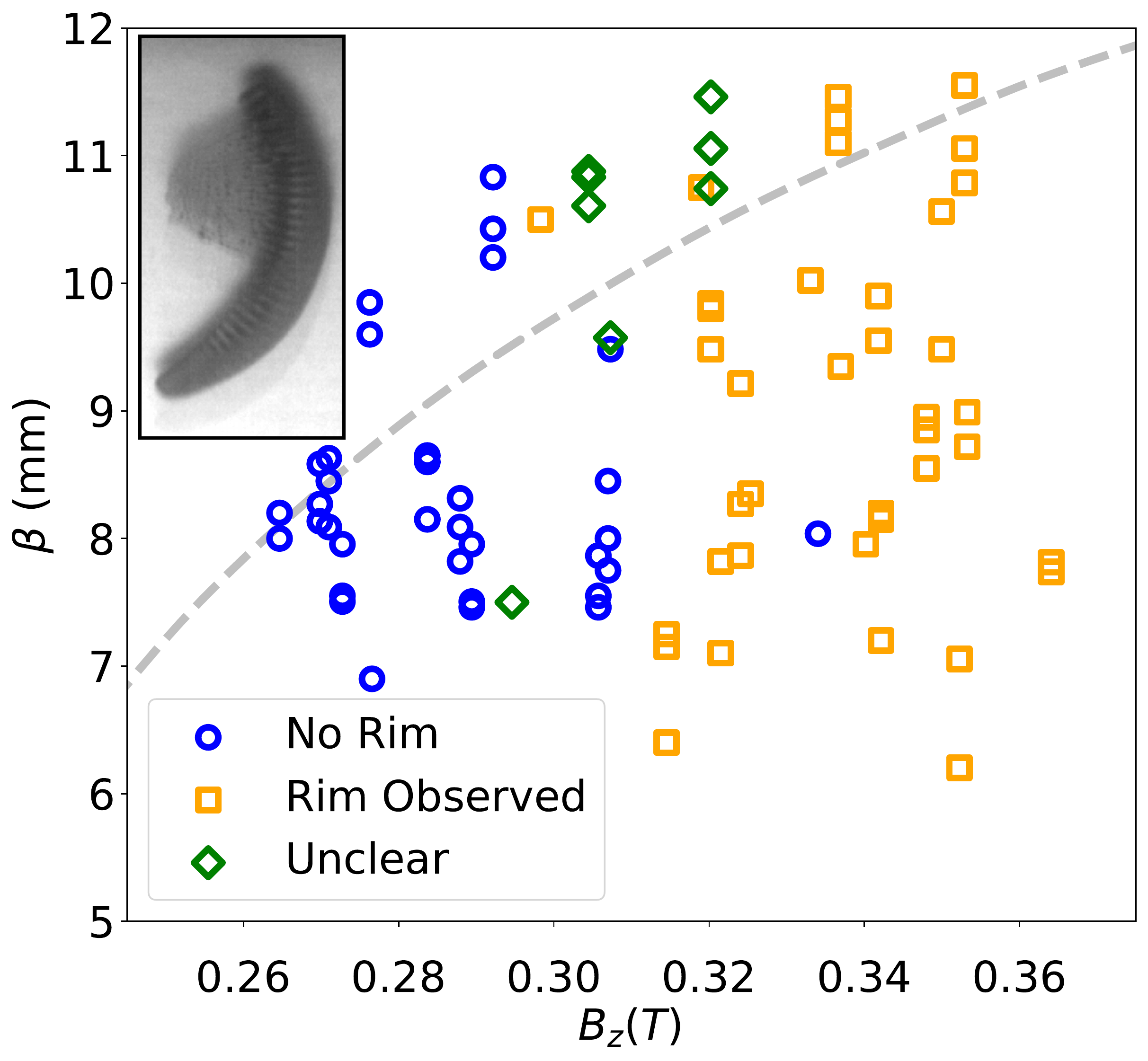}
    \caption{Rim formation as a function of $B_z$ and the normalized maximum spreading diameter. The dotted grey line corresponds to $f_R=0$ from Fig.~\ref{fig:rim_forces}. Above this line the calculated force acts radially outwards, and below the force acts inwards. Inset: a drop impact (We = 453, $V = 2.79 \mu$l, $v = 3.23$~m/s, $h_0 = 2.26$~mm, $B = 0.358$~T) pictured at $t=37$~s, showing partial rim formation.}
    \label{fig:rim_phase}
\end{figure}

To further probe rim formation and the effect of impact velocity, the set of droplet impact experiments carried out using a glass substrate for $2<h_0<5$~mm ($B_z\gtrsim0.25$~T) was studied further. These experiments are included in Fig.~\ref{fig:contact_line_phase}. Figure~\ref{fig:rim_phase} shows the occurrence of rim formation as a function of $B_z$ and the maximum spreading diameter. The droplet is considered to have formed a rim when there is a clear discrepancy between the height of the droplet edge and centre, as observed using both the top and side-view videos. Some cases are labelled ``unclear'' due to ongoing motion of the fluid at the end of the captured video.

Our initial hypothesis was that the transition between inwards and outwards radial force ($f_R=0$, Fig.~\ref{fig:rim_forces}) would cause rim formation. A ferrofluid element in a spreading droplet would be pushed towards that calculated radius, and reach dynamic stability there. In agreement with this explanation, the width of the rim (when observed) was generally thicker for smaller values of $B_z$ (see Supplementary Information). However, this hypothesis does not explain the onset of rim formation in Fig.~\ref{fig:rim_phase}, which is primarily a function of the magnetic field strength. It appears that above some threshold value ($B_z \gtrsim 0.31$~T), which does not depend on the extent of horizontal spreading, the vertical field is sufficient to pin the rim of the lamella in place as it slows during spreading. This pinning apparently overcomes the surface tension which acts to smooth the droplet and dissipate the rim. 

It is prudent to discuss some experimental uncertainties which could affect the conclusions here and may warrant further investigation. Firstly, the arbitrary limit on the recording time means that the results may not represent the equilibrium state of the rim. Secondly, many of these experiments produced less symmetric spreading than shown in Fig.~\ref{fig:rim_formation}, as the symmetry of the spreading was very sensitive to the position of the drop impact. When asymmetric impact occurred, a rim could form to one side of the impact point and spread along the shape of a circular rim (Fig.~\ref{fig:rim_phase}, inset). 


\begin{figure}
    \centering
    \includegraphics[width=8.6cm]{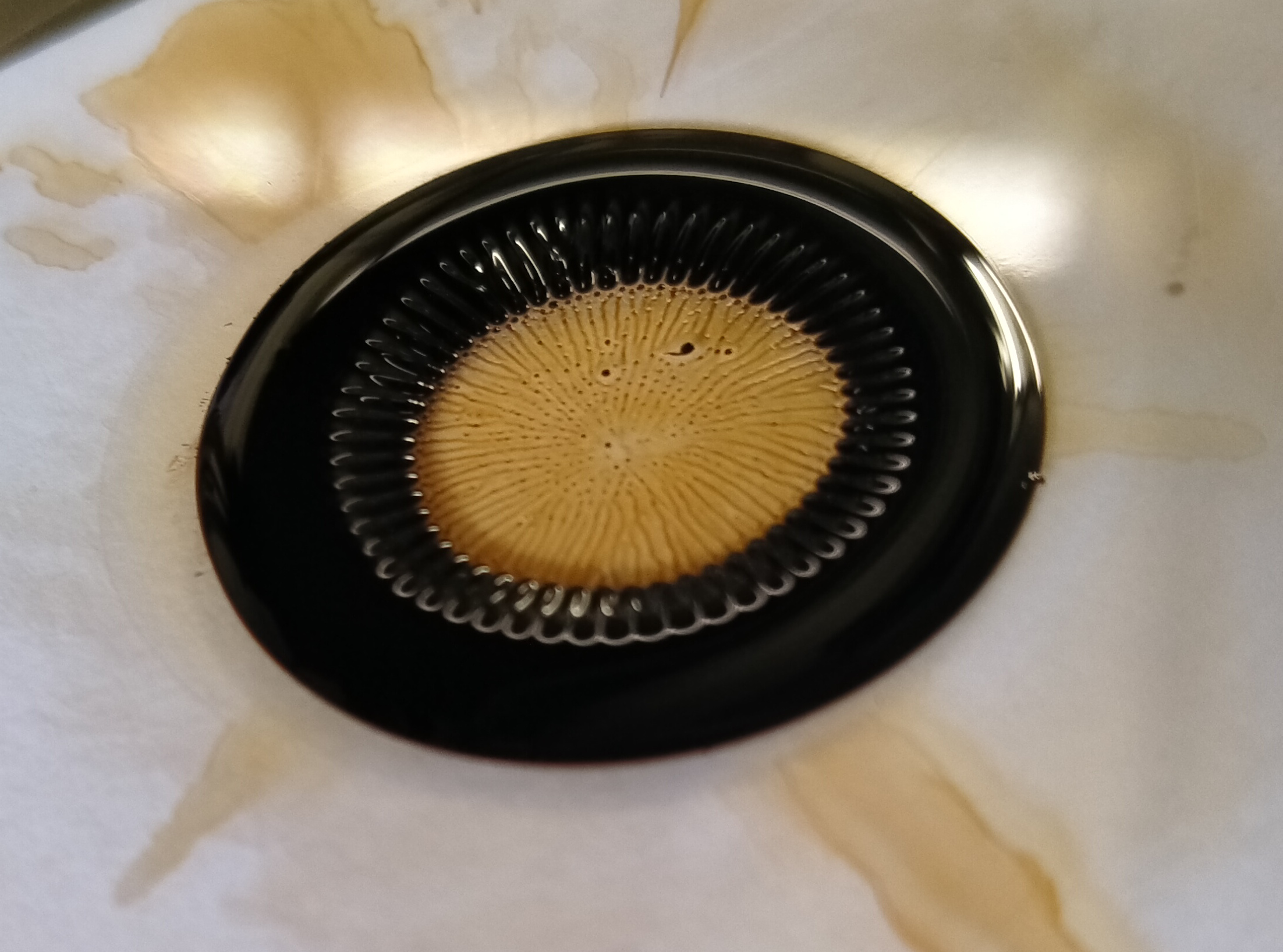}
    \caption{A rim formed following drop impact of a larger amount of ferrofluid than in Fig.~\ref{fig:rim_formation}.}
    \label{fig:rim_higher_volume}
\end{figure}

In a final experiment, a drop impact was carried out using a larger volume of ferrofluid, with the result shown in Fig.~\ref{fig:rim_higher_volume}. A rim was clearly formed, and as additional fluid was added to the stationary drop, the rim grew more in the inwards direction. The radial ridges that can be observed on the inner edge of the rim are likely to be related to those reported previously \cite{odenbach_colloidal_2009, Reimann_ridges}, in which lines rather than Rosensweig peaks appeared on the surface of a fluid when the magnetic field was tilted by 23\textdegree\space to the vertical. Other workers have observed a floating ferrofluid layer form a circular rim within a radially inhomogeneous field \cite{bushueva_evolution_2011}. These observations suggest that rim formation can be explained by the static magnetic field, but that a dynamic event such as a droplet impact is required to observe phenomena such as radially moving Rosensweig peaks, and as discussed earlier can affect the final droplet shape.

\section{Conclusion}

Ferrofluid drop impacts have been investigated within a non-uniform field of the kind readily produced by a simple permanent magnet. The work has focused on spreading droplets which form Rosensweig instabilities, and which do not splash or break up. In general, the outcomes were more dependent on the magnitude of the applied field than on the size, elongation, or velocity of the impacting drops as characterised by the Weber number. The non-uniform field is most simply characterized by the vertical component of the flux density at the impact surface, $B_z$.

The motions of the fluid-surface contact line and of the spreading droplet height were measured during the initial fast expansion to `maximum' spread ($\approx$5~ms after impact), and up to around 100~ms. At $B_z\approx0$, the drops retracted slightly after reaching maximum spread, and their height monotonically decreased. For $0<B_z\lesssim$~0.25~T peaks formed so that, after the initial fast spreading stage, the height of the spreading drop grew. The instabilities repelled each other so that the extent of the contact line also gradually increased. At $B_z\gtrsim$~0.25~T, the greater field produced more, smaller peaks so that there was monotonic height reduction and outwards spreading. 

The first and largest Rosensweig peaks were formed on the rim at the droplet edge, where there is a greater volume of fluid available than in the lamella. Although some motion and coalescence of peaks was observed over time, the largest peaks remained at the outer edge of the drop, with smaller peaks closer to the centre, for at least $\approx$1~s. This patterning is the inverse of what is observed for static pools of ferrofluid. Perturbations which evolved into the largest, outer peaks were apparent in the rim by the time of maximum spread ($\approx$5~ms). For relatively high fields, it was clear that these were instigated by the $B$-field. At lower fields, magnetic perturbations are slower to form and it is possible that Rosensweig peaks are primarily nucleated by crown-rim instabilities.

For higher fields ($B_z\gtrsim0.31$~T) the peaks at the outer edge of the spreading drop eventually coalesced to form a prominent, stationary rim near the edge of the magnet below. The transition to rim formation is primarily a function of the magnetic field strength, rather than being related to drop spreading. It is suggested that the rim is pinned by the magnetic field in a similar manner to previous observations of static ferrofluid drops.

The study of drop impacts generally has a broad parameter space, including (for example) the size, release height, and fluid properties of the drop. For ferrofluids this space has not been well studied, so that various interesting extensions to the current work are possible. For example, impacts which produce splashes can be studied, and the shape of the non-uniform $B$-field could be varied along with its magnitude. Impacts can also be studied for a wide range of surface properties, including different wettabilities, as well as random or designed surface structures (roughness) at any length scale. Ultimately there is the prospect of using drop impact to explore and control distributions of Rosensweig peaks, thereby producing interesting patterns that can be of practical interest.

\begin{acknowledgments}
The authors wish to thank Ekaterina Zossimova and Santhosh Kumar Pandian for minor contributions to this work.
\end{acknowledgments}

\bibliography{ferrofluids}

\end{document}